\documentclass{elsart}

\usepackage{graphics} 
\usepackage{amssymb}

\begin{document}

\begin{frontmatter}

\title{Overdamped thermal ratchets in one and more dimensions. Kinesin
transport and protein folding.}

\author{Ernesto Gonz\'alez-Candela} and
\ead{egonzaca@banxico.org.mx}
\author{V\'{\i}ctor Romero-Roch\'{\i}n\corauthref{cor}}
\corauth[cor]{Corresponding author.}
\ead{romero@fisica.unam.mx}

\address{Instituto de F\'{\i}sica. Universidad Nacional Aut\'onoma de M\'exico.
\\ Apartado Postal 20-364, 01000 M\'exico, D.F. Mexico.}

\begin{abstract}

The overdamped thermal ratchet driven by an external
 (Orstein-Uhlenbeck) noise is revisited. The ratchet we consider is
 unbounded in space and not necessarily periodic . We briefly discuss
 the conditions under which current is obtained by analyzing the
 corresponding Fokker-Planck equation and its lack of stationary
 states. Next, two examples in more than one dimension and related to
 biological systems are presented.  First, a two-dimensional model of
 a ``kinesin protein'' on a ``microtubule'' is analyzed and, second,
 we suggest that a ratchet mechanism may be behind the folding of
 proteins; the latter is elaborated with a multidimensional ratchet
 model.

\end{abstract}


\end{frontmatter}

\section{Introduction}

The ratchet mechanism to produce directional current or motion
has become a paradigm of the interplay of nonlinear phenomena with
Brownian motion and external noise in mesoscopic 
systems\cite{Magnasco,Reimann,Hanggi1}.
We shall use the term ``ratchet'' for any conservative mechanical 
system that has an ingrained spatial asymmetry; a one-dimensional
example is a particle in a saw-tooth like potential. Due to the potentiality
of these systems to describe mesoscopic systems and
their processes, such as biological ones\cite{Astumian1}, 
one should consider the ratchet 
to be immersed in a thermal bath, giving rise to Brownian motion.
This adds dissipative and stochastic thermal forces that are
related to each other 
through the fluctuation-dissipation theorem\cite{vanKampen,Risken}.
As it was clearly pointed out by Feynman in his 
{\it Lectures}\cite{Feynman}, the Second Law of Thermodynamics
implies that a ratchet under the above conditions
cannot generate directional motion. It is by now very well 
established that in order to obtain current or directional
motion, it is indispensable to have the presence of an
{\it external} unbiased time dependent force. This force can
be deterministic, a so-called ``rocking'' ratchet, see Ref.\cite{Magnasco},
or stochastic in nature, see Ref.\cite{Doering}. In any case, 
any of those forces should be of zero average
in time in order not to produce any biased current; in the
present article we shall be concerned with external stochastic
forces only. 

It is the opinion of the authors that the origin
of the appearance of the current is not fully understood
yet, although it is clear what the conditions are for 
its existence. In particular, 
a lot of attention has been devoted to the case in which the
ratchet variable is periodic, such as an angle $\theta \in [0,2\pi]$, 
and many analytic and numerical results have been 
obtained\cite{Reimann,Bartussek}, yielding a clear picture of how current
occurs in a stationary state. We believe the
studies are not as exhaustive for the cases where the ratchet 
variable is not periodic, namely when the particle can move in
all space, namely, when the position of the particle can take all
real values, $x \in (-\infty,\infty)$.
In this case, even in the absence of external forces, the
system does not reach an stationary state\cite{vanKampen}. 
One of the purposes of this article is to add to the understanding
of the appearance of the current in the latter situation, 
by studying a novel {\it non-periodic} one-dimensional overdamped
ratchet; this is done in Section 2. 

Having established the conditions for the production 
of current in one dimension, we then proceed to the second
purpose of the article and extend the
model to more than one dimension in order to study two 
biology-related problems. One is the case of the motion
of a kinesin protein on a microtubule, in Section 3, and the other is the 
process of protein folding in Section 4. The first has been studied 
already by
several authors\cite{Astumian,Mateos} but here we consider a
protein in two dimensions and the microtubule as a quasi-one dimensional
structure. Our model of protein folding is at present 
speculative, and the idea is centered on the hypothesis that
the energy landscape of the protein has a ratchet-like structure;
by further assuming that there is an {\it external} agent
that consumes energy, which could be the presence of chaperone
proteins\cite{Alberts}, one has all the ingredients to obtain
directed motion. We shall show that, understanding by protein folding
the search and finding of a target point or small section in the
energy landscape, the ratchet mechanism is extremely fast with an
almost perfect efficiency.
An interesting consequence of this process is that the landscape need
not have a funnel structure\cite{Wolynes,Dill}.

\section{Current in a one-dimensional overdamped ratchet}

In this section we revisit the one-dimensional overdamped ratchet system,
considering both periodic and non-periodic potentials.
The main purpose
here is to argue that the appearance of a current is a {\it generic} 
property of bounded asymmetric conservative systems, in interaction 
with a thermal bath, due to the presence of external
(stochastic or deterministic) non-biased time dependent forces. 
That is, by showing that the absence of current
without the external force must follow from the Second Law, 
we argue that there is nothing to
prevent a current once the conditions required by the 
Second Law have been relaxed. The details of the numerical
techniques we use can be found in Refs.\cite{RR1,RR2}.

\subsection{No current in the absence of external 
time-dependent forces} 

\begin{figure}[ht]
\centering
\scalebox{.5}{\includegraphics{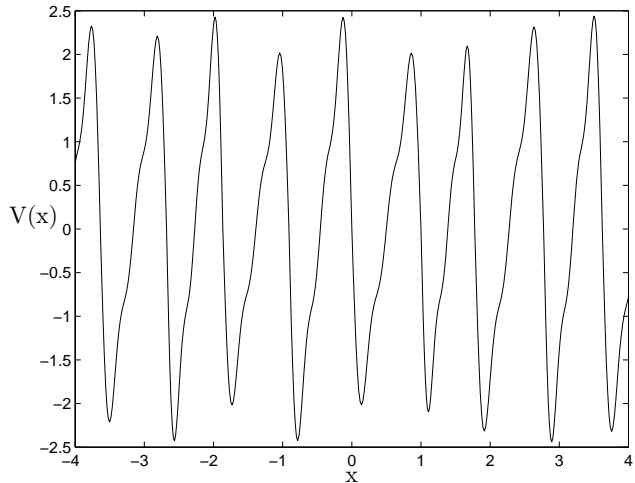}}
\caption{Ratchet potential with quenched random disorder both
in the potential barriers and in the spatial periods.
See Eq.(\ref{ratchet}).}
\label{pot-ran}
\end{figure}

We initiate by writing down the 
the Langevin equation in the overdamped case for a particle 
subjected to a conservative potential $V(x)$, without 
an external time dependent force,
\begin{equation}
\gamma {dx \over dt} = - {\partial V \over \partial x} + f(t). 
\label{Langevin}
\end{equation}
We consider a ratchet potential with a ``quenched'' random disorder
both in the potential barriers and in the spatial periods, namely,
\begin{eqnarray}
V(x) &=& - V_0^{(i)} \left[ \sin \left(\frac{2 \pi 
(x-y_i)}{\lambda_i}\right)
+ \frac{2}{5} \sin \left(\frac{4 \pi (x-y_i)}{\lambda_i}\right) +
\frac{1}{10} \sin \left(\frac{6 \pi (x-y_i)}{\lambda_i}\right)
\right] \nonumber \\
&&  \>\>\>\>{\rm if} \>\>\>y_{i} \le x < y_{i+1} ,\label{ratchet}
\end{eqnarray}
with $i$ taking all integer values and where the positions $y_i$ are chosen 
at random; $y_0$ is arbitrarily
set equal to 0. Clearly, the spatial periods are given 
by $\lambda_i = y_{i+1} - y_i$. A typical form of the ratchet potential 
$V(x)$ is 
shown in Fig. \ref{pot-ran}.  Note that this potential
is not periodic. The usual {\it periodic}
ratchet potential is $V_0^{(i)} \equiv V_0$ and $\lambda_i \equiv 
\lambda$ for all $i$\cite{Reimann}.

Returning to the Langevin equation (\ref{Langevin}),
$\gamma$ is the friction coefficient 
and $f(t)$ is the thermal force exerted by the thermal bath. The
stochastic properties of this force are 
\begin{equation}
\langle f(t) \rangle = 0 \>\>\>\>{\rm and} \>\>\>\>
\langle f(t) f(t^\prime) \rangle = 2 \gamma kT \delta(t - t^\prime) 
.\label{fT}
\end{equation}
It is further assumed that this process is Gaussian, 
so that Eqs.(\ref{fT}) suffice to 
completely specify the process $f(t)$. 
In the second equation $T$ is the temperature of
the bath and the expression for the correlation
of the force is the 
Fluctuation-Dissipation relation for this problem; it 
imposes detailed balance\cite{vanKampen,Risken}.

Since the process $x(t)$ representing the position of 
the particle is Markovian, then, knowledge of the
conditional probability distribution $\rho(x,t | x_0)$ 
is enough to determine the process. It can 
be shown that this function obeys the following 
Fokker-Planck equation\cite{vanKampen},
\begin{equation}
{\partial \rho(x,t) \over \partial t } = 
{1 \over \gamma} {\partial \over \partial x}
\left({\partial V \over dx} + kT {\partial \over \partial x}\right) 
\rho(x,t) , \label{FP}
\end{equation}
supplemented by the initial condition $\rho(x,0|x_0) = \delta(x-x_0)$.

In this discussion we are considering that the 
particle can move in all space, namely, $-\infty < x <
\infty$. Thus, the normalization of the distribution,
\begin{equation}
\int_{-\infty}^{\infty} \rho(x,t) \> dx = 1,  \label{norma}
\end{equation}
imposes the boundary conditions
\begin{equation}
\rho(x,t) \to 0 \>\>\>\> {\rm for}\>\>\>\> x \to \pm \infty .\label{BC}
 \end{equation}

The Fokker-Planck equation, Eq.(\ref{FP}),  
is a continuity equation for the probability distribution. 
Therefore, one can read off the probability density current, 
\begin{equation}
j(x,t) = -{1 \over \gamma}{\partial V \over \partial x} \rho(x,t) -  
{kT \over \gamma} {\partial \rho(x,t) \over \partial x} . \label{j}
\end{equation}
We may define the ``total" current as,
\begin{eqnarray}
J(t) &=& \int_{-\infty}^{\infty} j(x,t) \> dx \nonumber \\
&=& -{1 \over \gamma}
\langle {\partial V \over \partial x} \rangle  \nonumber \\
&=&  {d \over dt} \langle x(t) \rangle , \label{J}
\end{eqnarray}
where the last two lines follow from the definition 
of the density current and the Fokker-Planck equation, 
Eqs.(\ref{j}) and (\ref{FP}).

If there exists a stationary distribution, $\rho_s(x)$, 
then detailed balance means that the stationary density 
current vanishes, $j_s(x) = 0$, and therefore that the 
stationary distribution is
\begin{equation}
\rho_s \sim e^{- \beta V(x)} , \label{rhos}
\end{equation}
with $\beta = 1/kT$.  Further, if $\rho(x,t)$ is not 
the stationary distribution,
then, detailed balance implies that as $t \to \infty$, 
such a distribution approaches the stationary one.
This is equivalent to the $H$-theorem. 

In the context of ratchet-like potentials, one 
is faced with potentials that are bounded, that is,
$V_1 \le V(x) \le V_2$, for all $x$. Thus, strictly 
speaking, a stationary distribution cannot be 
reached\cite{vanKampen}. However, it is clear 
that stationarity is achieved in the following sense,
\begin{equation}
\lim_{t \to \infty}^{} 
{\rho(x,t) \over \rho(x^\prime, t)} = 
e^{-\beta (V(x) - V(x^\prime))} .\label{station}
\end{equation}

\begin{figure}[ht]
\centering
\scalebox{.5}{\includegraphics{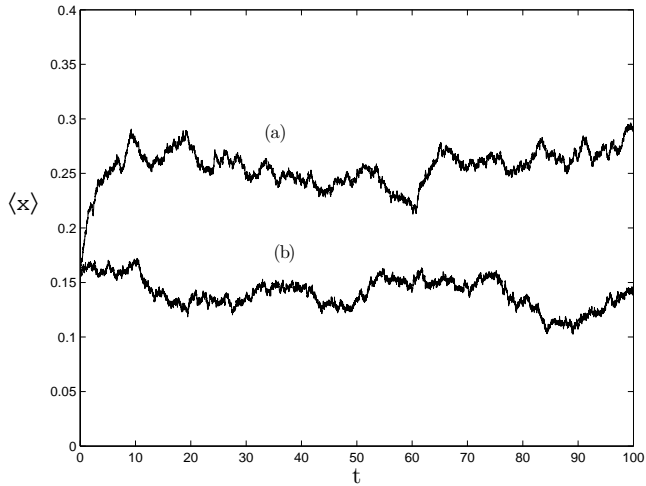}}
\caption{Average position of the particle $\langle x(t) \rangle$
as a function of time, for the case of no external 
time-dependent force, see Eq.(\ref{Langevin}). (a) For
a disordered ratchet potential. (b) For a periodic
ratchet potential. In both cases the initial condition
is $x = - 0.04$. Note that the particle, on the average,
never leaves the well where it started, see Fig. 
\ref{pot-ran}.}
\label{prom}
\end{figure}

Therefore, it must be true  that, in the above sense, 
the probability density current $j(x,t)$ and the total 
current $J(t)$ must vanish as $t  \to \infty$. However, 
this does not suffice to prevent an arbitrary current 
or displacement of the particle, as is routinely asserted 
in essentially all articles dealing with this problem. 
The main point that we want to emphasize here is that, 
since detailed balance is in accordance with
 the Second Law of Thermodynamics,
the total current $J(t)$ must be bounded for all times. 
Using Eq.(\ref{J}), this can be more precisely written
as the following requirement,
\begin{equation}
| \langle x(t) \rangle - x_0 | = \left| \int_0^t \> 
d\tau \> J(\tau) \right| < \bar \lambda (x_0) , \label{bound}
\end{equation}
where $x_0$ is the initial position of the particle 
and $\bar \lambda (x_0)$ is the distance 
between the adjacent local maxima of the potential 
where the particle was initially at $x = x_0$. 
We have exhaustively numerically verified that this 
holds for ratchet potentials such as those ones in 
Fig. \ref{pot-ran}, and the results are exemplified in 
Fig. \ref{prom}. 

For ratchet potentials this means that {\it on the average}
the particle cannot leave the local well where it was
initially; if it did, nothing
would prevent the particle to ``jump" to another well, 
and so on, thus moving an arbitrary distance. 
In other words, Eq.(\ref{bound}) expresses that 
fact that it is not possible to obtain a current that, 
on the average, could generate motion that would 
transport the particle farther than its initial well. 
Otherwise, this would yield the possibility to perform work 
on some external load or agent\cite{Parrondo} 
violating Kelvin's statement of the Second Law: 
``A transformation whose only final result is to
transform into work heat extracted from a source
which is at the same temperature 
throughout is impossible"\cite{Fermi}. In the 
opinion of the
authors, this is what Feynman, in his
{\it Lectures}\cite{Feynman}, wants to convey 
in his discussion of a ratchet engine.

\begin{figure}[ht]
\centering
\scalebox{.5}{\includegraphics{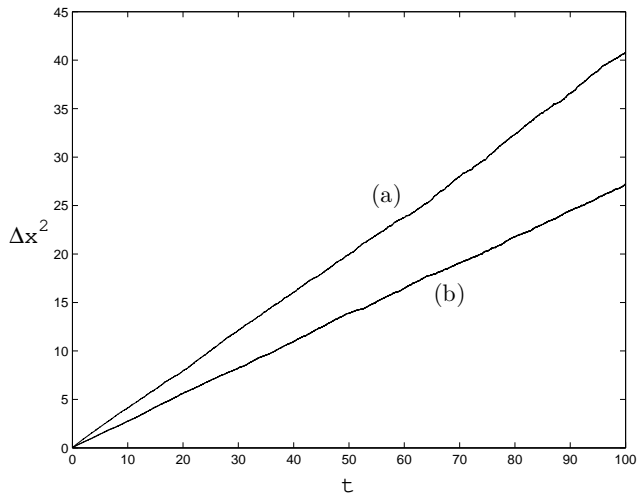}}
\caption{Second moment of the particle position
distribution $\Delta x^2 = \langle x^2 (t)\rangle - \langle x(t) \rangle^2$ 
as a function of time, for the case of no external 
time-dependent force, see Eq.(\ref{Langevin}). (a) For
a disordered ratchet potential. (b) For a periodic
ratchet potential. As discussed in the text 
the particle performs normal diffusion.}
\label{seg}
\end{figure}

It is important to stress that we consider
high enough temperatures so that the particle, even though it does
not move on the average, it does perform normal diffusion.
This is shown in Fig. \ref{seg}.

\subsection{Current in the presence of an external 
time-dependent force}

In the presence of a time-dependent external force 
$F(t)$, the Langevin equation reads,
\begin{equation}
\gamma {dx \over dt} = - {\partial V \over \partial x} 
+ f(t) + F(t) ,\label{Lang2}
\end{equation}
where the thermal force $f(t)$ obeys the same 
properties as before, see Eqs.(\ref{fT}). 
As an example, the external force $F(t)$ may be generated 
by an Orstein-Uhlenbeck process
\begin{equation}
\tau_0 {dF \over dt} = - F + \zeta(t) ,\label{OU}
\end{equation}
with $\zeta(t)$ a given Gaussian, white, stochastic process:
\begin{equation}
\langle \zeta(t) \rangle = 0 \>\>\>\> \langle \zeta(t) 
\zeta(t^\prime) \rangle = f_0^2 \delta(t - t^\prime).
\label{zt}
\end{equation}
The parameters $\tau_0$ and $f_0$ determine the process. 
The first one is the correlation time
for $F(t)$ and $f_0$ is a measure of its strength. 
In order to avoid any bias by this force, one must 
consider the initial condition $F(0) = 0$.

Equations  (\ref{Lang2}), (\ref{fT}), (\ref{OU}) and 
(\ref{zt}) are equivalent to the following
bivariate Fokker-Planck equation\cite{Bartussek},
\begin{eqnarray}
{\partial W(x,F,t) \over \partial t }& =& 
{1 \over \gamma} {\partial \over \partial x}
\left(- F + {\partial V \over \partial x} + 
kT {\partial \over \partial x}\right) 
W(x,F,t) \nonumber  \\
&&+ {1 \over \tau_0} {\partial \over \partial F}
\left(F + {f_0^2 \over 2 \tau_0} {\partial \over \partial x}\right) 
W(x,F,t) , \label{FP2} 
\end{eqnarray}
where $W(x,F,t|x_0,0)$ is the conditional probability 
distribution to find values $x$ and $F$ for
the corresponding stochastic variables, given that at 
$t= 0$, $x = x_0$ and $F=0$. Initially, it is
$W(x,F,0|x_0,0) = \delta(x - x_0) \delta(F)$.

By simple inspection we find that, even if a stationary 
distribution 
$W_s(x,F)$ exists, equation (\ref{FP2}) does
not obey detailed balance. The ``offending" term is
\begin{equation}
{1 \over \gamma} {\partial \over \partial x}(-F W(x,F,t)) 
.\label{offending}
\end{equation}
This, of course, only means that $F(t)$ is ``external". 
Namely, $F$ acts on $x$ but not the opposite,
Thus, there is no mechanism to establish equilibrium among 
the particle, the thermal bath  
and the external agent. Therefore, the system can withdraw 
energy from the external source and
generate motion. In other words, there is nothing to prevent 
the total current from taking any value different from zero:
\begin{equation}
J(t) = -{1 \over \gamma}  \int_{-\infty}^{\infty} dx \int_{-\infty}^{\infty}
dF \> W(x,F,t)  {\partial V (x) \over \partial x} \ne 0 ,\label{J2}
\end{equation}
or
\begin{equation}
J(t) = {d \over dt} \langle x(t) \rangle \ne 0 .
\end{equation}
In other words, in this case the Second Law does not 
impose any restriction on the appearance of a current.
The Second Law will now impose restrictions on the {\it
efficiency} of the process but that is not discussed here.
Of course, if the potential $V(x)$ 
does have a ``left-right" symmetry, the current vanishes 
as one should expect. 

\begin{figure}[ht]
\centering
\scalebox{.5}{\includegraphics{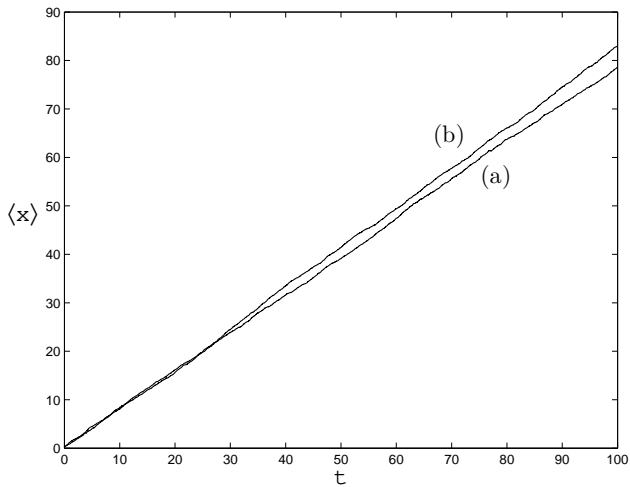}}
\caption{Average position $\langle x(t) \rangle$ as a function
of time in the presence of an  external 
time-dependent force. (a) For
a disordered ratchet potential. (b) For a periodic
ratchet potential.}
\label{current}
\end{figure}

In Fig. \ref{current} we exemplify the appearance of a current, both
for the disordered potential and for the periodic one.
On top of the current, the particle also performs normal
diffusion, i.e. for long times $<x^2(t)> - <x(t)>^2 \sim D t$.

\section{A two-dimensional ``kinesin on a microtubule'' model}

During the last few years there has been an increasing interest in
studying the statistical behavior of the transport phenomena inside
the cell carried out by protein motors and it has been proposed by
several authors
that ratchet models may be relevant in the description of 
these
processes\cite{Magnasco,Reimann,Doering,Bartussek,Astumian,Ajdari}. 
Protein motors are responsible for
carrying diverse kind of vesicles from one site
to another in the cell in a much more efficient way than the 
obtained by simple diffusion\cite{Alberts}. To accomplish this 
task they consume chemical energy, usually stored in the form of
adenosine-triphosphate (ATP), to convert it into mechanical motion
\cite{Reimann}.  Due to their dimensions, the erratic collisions with
the solvent molecules represent a non negligible contribution to the 
protein motors
dynamics and appear in the form of friction and thermal stochastic
forces. This is why they are also called Brownian motors or
molecular motors\cite{Astumian1}.

Molecular motors, as kinesin and dynein, present directed motion only
when they are attached to microtubules, otherwise they perform standard 
Brownian motion. It has been also observed that
they do not hydrolyze ATP at an appreciable rate unless they
are attached\cite{Reimann}. Experimental observations of kinesin and 
dynein motion on microtubules\cite{Wang} have shown that kinesins 
move mainly along one single protofilament, while dyneins visit often 
several protofilaments. One important
characteristic of the microtubules is that they have an intrinsically
periodic and asymmetric structure\cite{Alberts}.

Most of the models motivated by these experimental results were
initially restricted 
to the one dimensional case but it has become clear that 
more dimensions are 
needed\cite{Mateos,Cilla,Zhao,Middleton,Kostur,Bao,Bao2,Eichhorn,Lipowsky,Nieu}. 
Some of these works\cite{Lipowsky,Nieu} approach the problem imposing 
{\it ad hoc} asymmetric probabilities thus obtaining transport.  The 
others use Langevin equations and, in particular, those
in Refs.\cite{Mateos,Cilla,Zhao,Middleton}  
have introduced more detailed models to understand the particular behavior of 
kinesin in microtubules.

As an step forward in the description of these processes, we are
considering here a model for a kinesin on a microtubule as a ratchet
in a two dimensional space, immersed in a thermal dissipative bath,
and subjected to an external force of zero mean varying stochastically
in time. At present we study general statistical properties of the
model; a more detailed study and its comparison with real kinesin
proteins will be presented elsewhere. Nevertheless, the
results shown here resemble qualitatively what is observed in actual
experiments with motor proteins and microtubules\cite{Alberts}.

\begin{figure}[ht]
\centering
\scalebox{.75}{\includegraphics{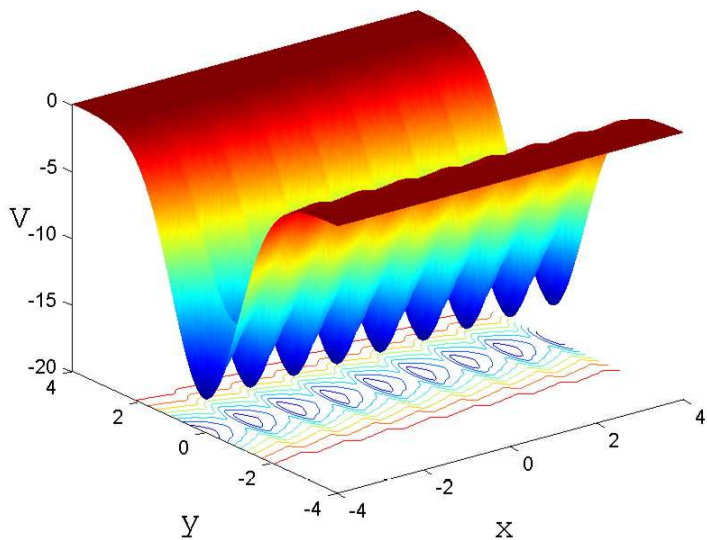}}
\caption{Two-dimensional potential with an infinite attractive filament
(or channel) along the $x$ coordinate and a ratchet structure inside
of it. The attractive filament has a mean depth of $V_1=15$ and a width
determined by $\sigma=2$, while $\lambda=1$ represents the period of
the ratchet and $V_0=2$ its amplitude. The longitudinal projection of
this potential is similar as the used in one dimension models. In the
transversal direction, $-\infty < y < \infty$. }
\label{potcan}
\end{figure}

In order to model the interaction forces between the motor protein
and the microtubule a two-dimensional potential is proposed, whose
shape resembles a long attractive filament (or channel) with a typical
ratchet structure inside on the longitudinal direction, as shown in
Fig. \ref{potcan}.
The potential $V(x,y)$, called for simplicity the filament or microtubule
potential, is given by the following function
\begin{equation}
V(x,y) = - e^{-y^2/\sigma^2}\left[V_1 + 
V_0\left(\sin\left(\frac{2\pi x}{\lambda}\right)
+\frac{4}{10}\sin\left(\frac{4\pi x}{\lambda}\right)
+\frac{2}{10}\sin\left(\frac{6\pi x}{\lambda}\right)\right)\right],
\label{potfil}
\end{equation}
where $V_1$ is the mean depth of the attractive filament and its width
is determined by $\sigma$; $\lambda$ represents the period of the ratchet
and $V_0$ the amplitude.  The system is immersed in a thermal bath at 
temperature $T$ and there is present an external random force with zero 
mean that represents in the model the consumption of ATP by 
the protein motor. Depending on the temperature of the bath and on
the statistical properties of the external force, the
particle may be inside or outside the microtubule, 
i.e. $|y| \le \sigma$ or $|y| > \sigma$ . Inside, it will 
feel the effect of the ratchet potential and may generate 
directed motion; outside, it will
perform free Brownian motion. 

The dynamics of the model are represented by the following 
coupled Langevin equations for a Brownian particle moving 
in an $x-y$ coordinate system: 
\begin{equation}
\gamma\frac{dx}{dt} = -\frac{\partial V(x,y)}{\partial x} 
+ f_{x}(t) + F_{x}(t),
\label{elx}
\end{equation}
and
\begin{equation}
\gamma\frac{dy}{dt} =  -\frac{\partial V(x,y)}{\partial y}  
+ f_{y}(t) + F_{y}(t),
\label{ely}
\end{equation}
where $\gamma$ is the friction coefficient and $f_{i}(t)$, 
with $i=x,y$, is the force exerted by the thermal bath. The stochastic 
properties of this force are
\begin{equation}
\langle f_i(t)\rangle=0 \>\>\>\>{\rm and}\>\>\>\>
\langle f_i (t)f_j (t') \rangle = 2\gamma kT\delta (t-t') \delta_{ij}.
\end{equation}
It is further assumed that these processes are Gaussian. 
The external force is
$F_{i}(t)$ with $i=x,y$. Due to the lack of information 
with respect to 
the actual statistical properties of ATP consumption by the kinesin, 
and which is one 
the main difficulties to make precise predictions, 
we have decided to employ a stochastic force
generated by an Ornstein-Uhlenbeck process
\begin{equation}
\tau_0\frac{dF_{i}(t)}{dt} = -F_{i}(t) + \zeta_i(t),
\end{equation}
with $\zeta_i(t)$ a given Gaussian, white, stochastic process
\begin{equation}
\langle \zeta_i(t)\rangle = 0 \>\>\>\>{\rm and}\>\>\>\> \langle
\zeta_i(t)\zeta_j(t')\rangle = f_o^2
\delta (t-t') \delta_{ij}.
\end{equation}
The parameters $\tau_0$ and $f_0$ determine the process and represent
the correlation time and the magnitude of the external force,
respectively. As it can be observed, the relations for 
$f_i(t)$ and $F_i(t)$ are
exactly the same for both directions and they are ruled by the same
parameters. In consequence, they act isotropically. 
However, each of them arise from an 
independent Gaussian noise, and thus, there is no 
correlation among them.

We now present the main 
statistical properties of this model. In all the results
presented here the initial conditions are $x(0)=y(0)=0$. Variations of
the different parameters, a thorough description of the behavior of
the first and second moments of the position distribution of the
particle, as well as a comparison with actual motor proteins, cannot
be presented here due to the briefness of this report and will be
discussed elsewhere.

\begin{figure}[ht]
\centering
\scalebox{.50}{\includegraphics{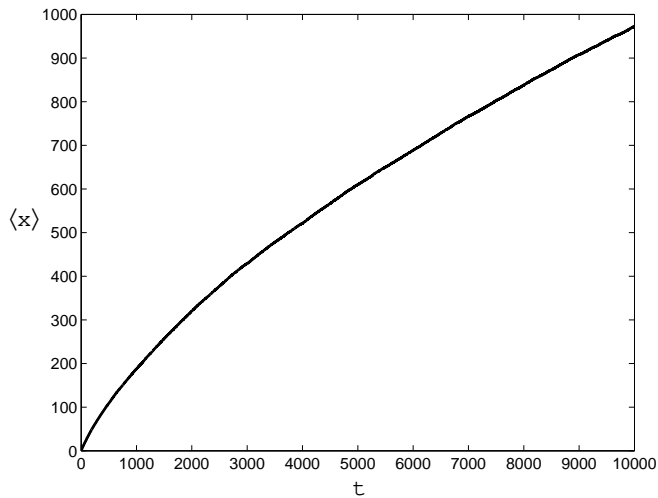}}
\caption{Average $x$-position of the particle as a function of time.}
\label{x-cur}
\end{figure}

\begin{figure}[ht]
\centering
\scalebox{.25}{\includegraphics{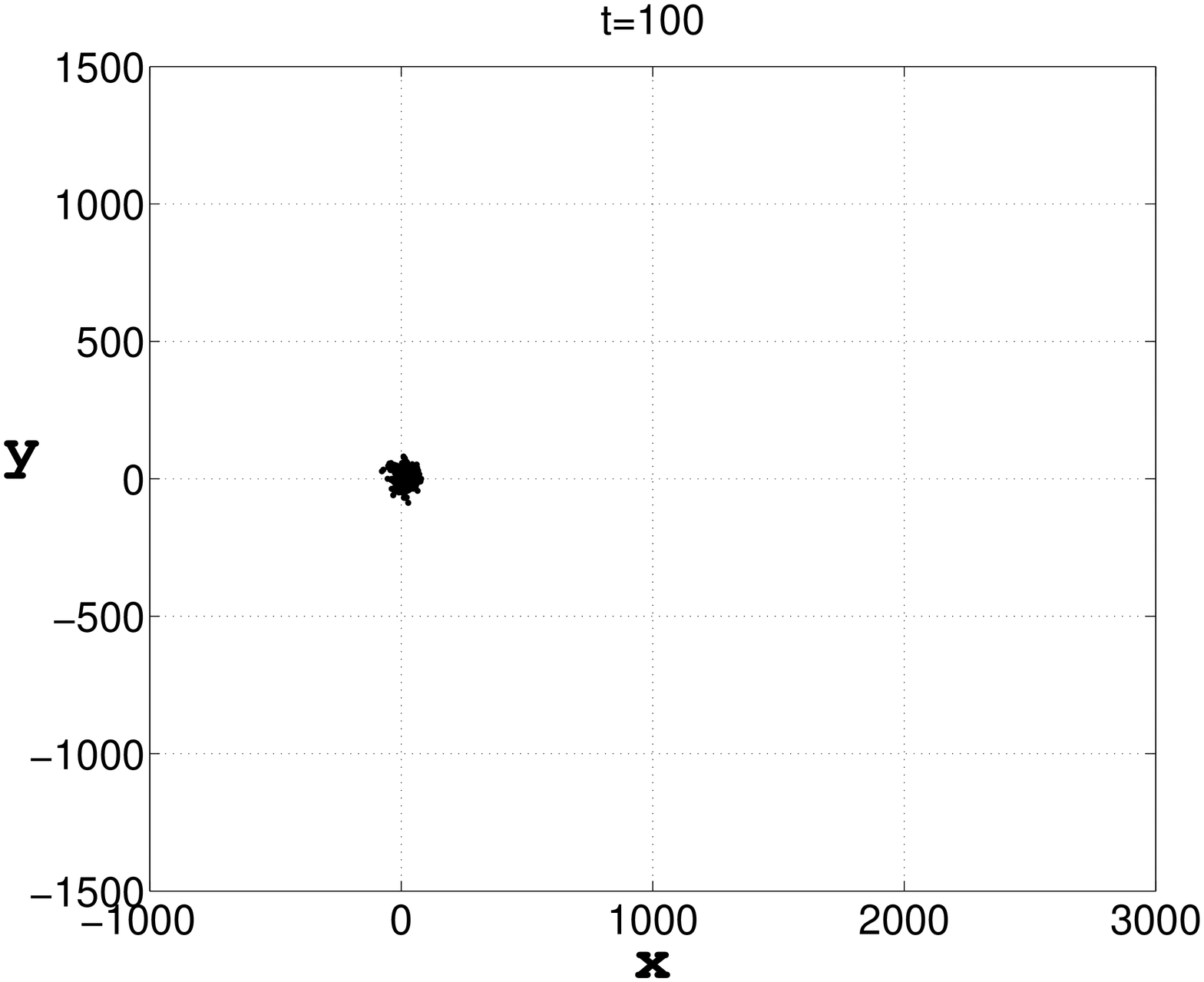}}
\scalebox{.25}{\includegraphics{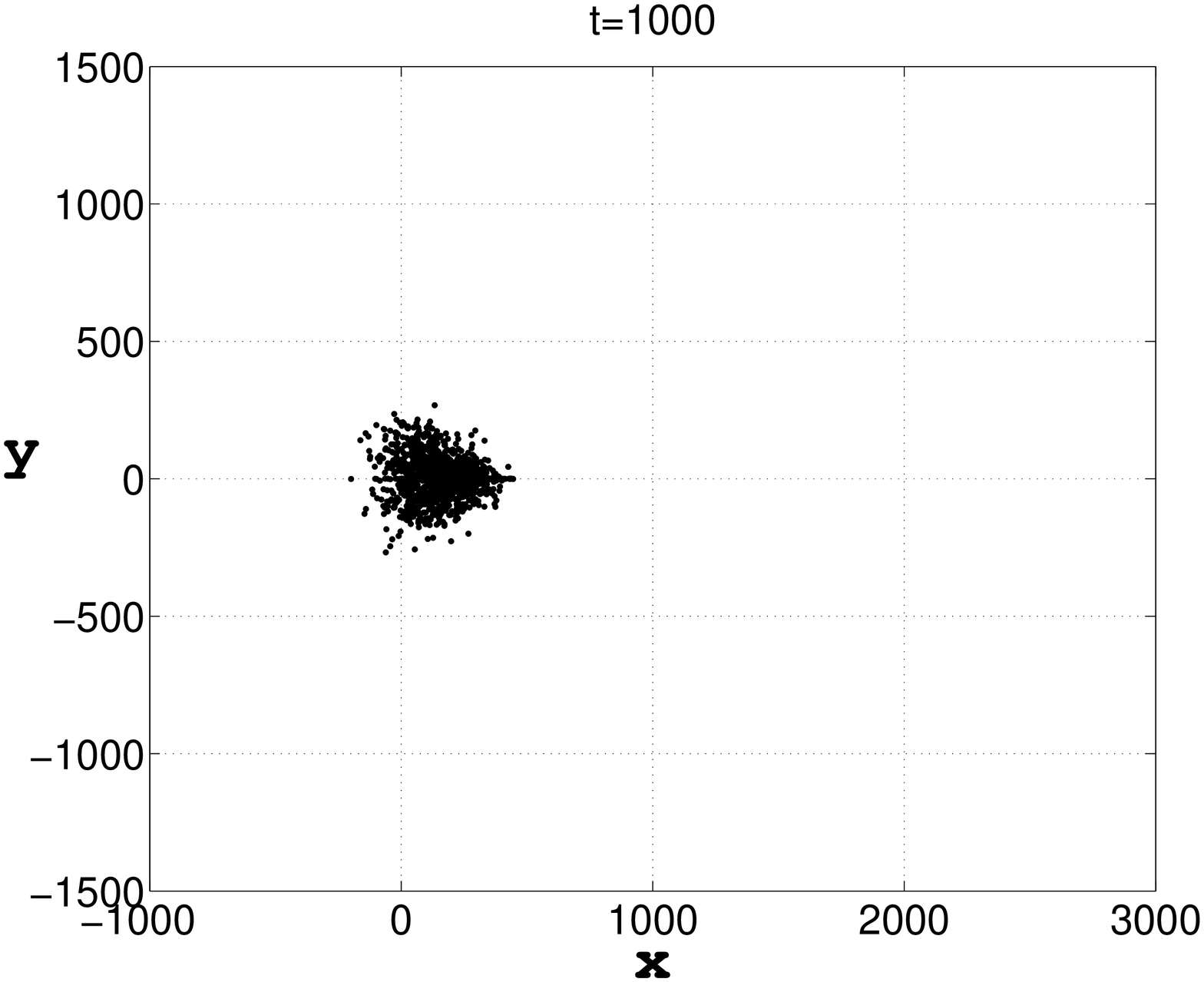}}
\scalebox{.25}{\includegraphics{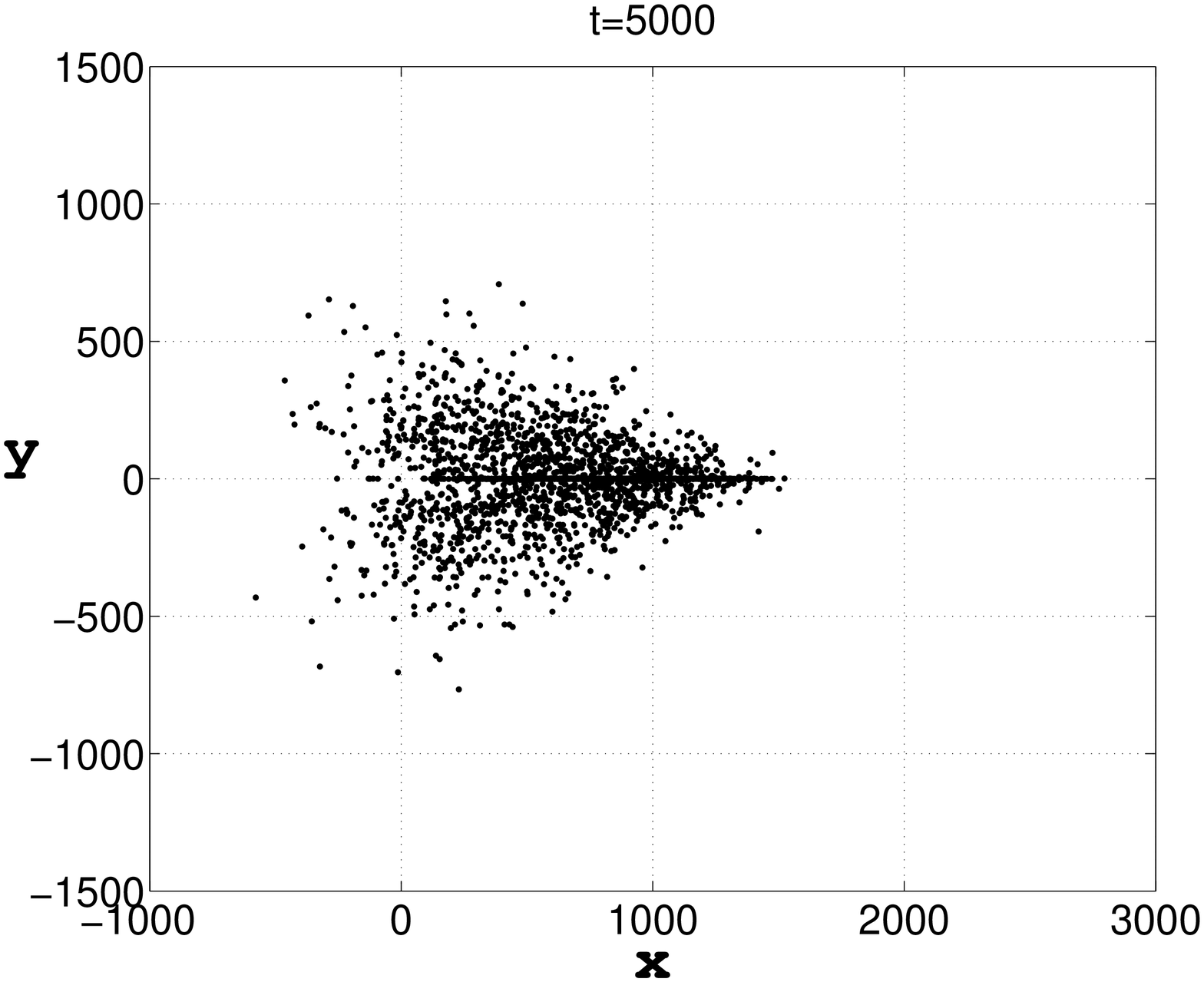}}
\scalebox{.25}{\includegraphics{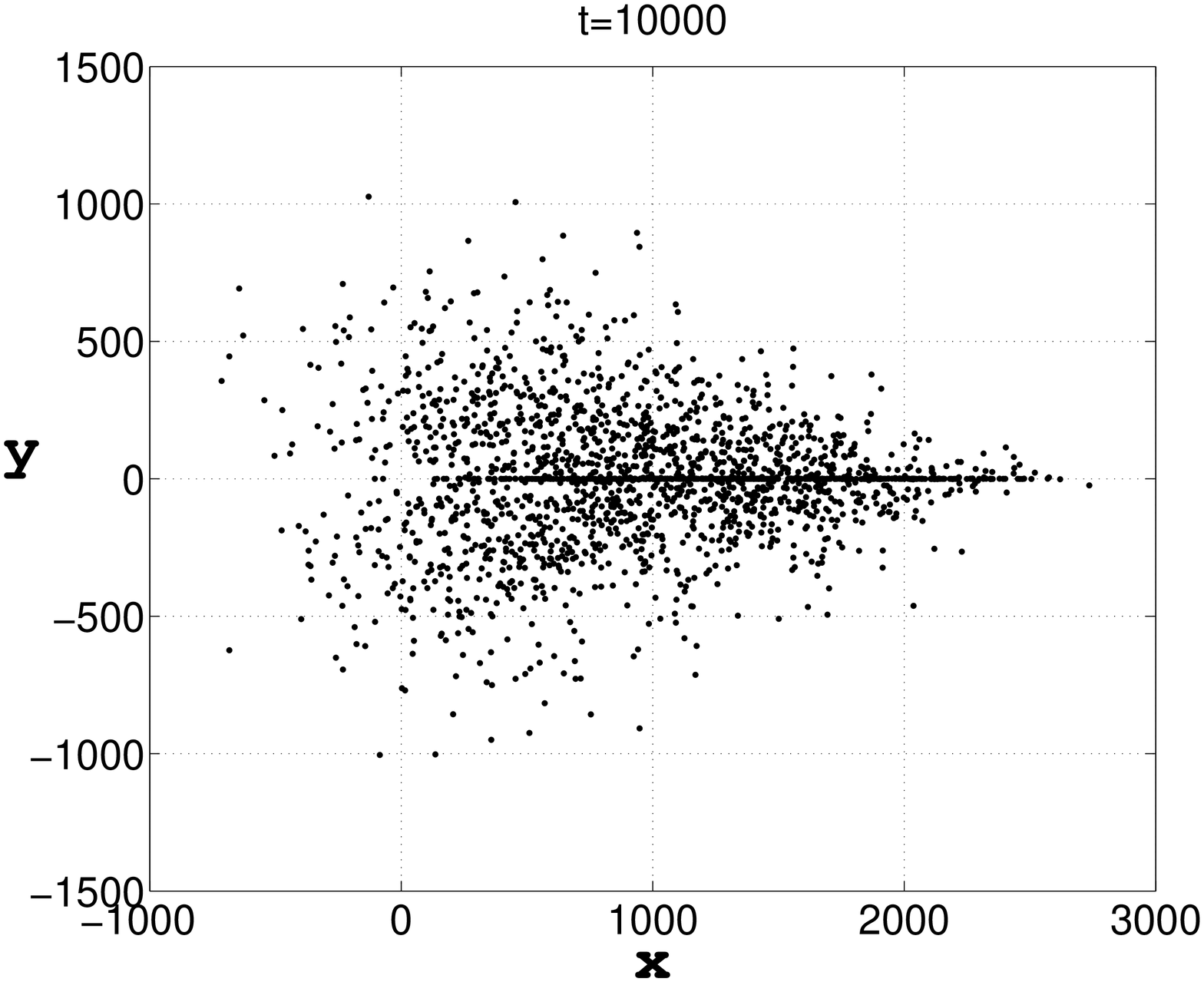}}
\caption{Snapshots of the position probability distribution
$\rho(x,y,t)$ for four different times, obtained from 2000
realizations of the process.}
\label{nubes}
\end{figure}

\begin{figure}[ht]
\centering
\scalebox{.50}{\includegraphics{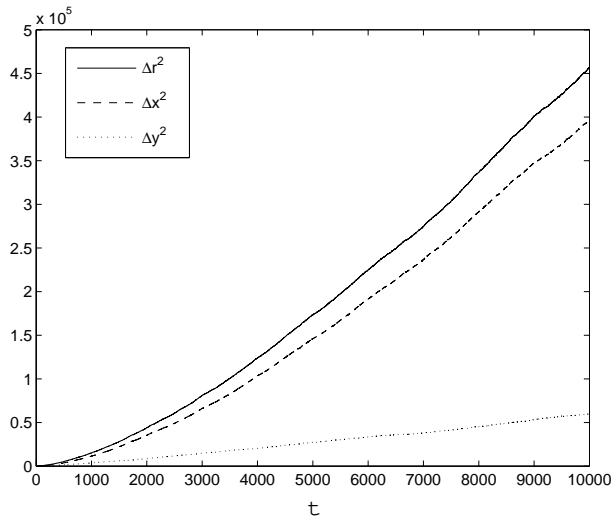}}
\caption{Total second moment $\Delta r^2$ and its components in $x$ and
$y$, see Eq.(\ref{Jxy}), as a function of time. Note that both $\Delta r^2$
and $\Delta x^2$ show superdiffusive behavior, while $\Delta y^2$ shows
normal diffusion.}
\label{msd}
\end{figure}

As one should expect, the presence of the external time dependent force
produces a net current along the $x$ direction due to the asymmetric
properties of the ratchet inside the filament. In the present case,
the total current is a two-dimensional vector,
\begin{equation}
\vec J(t) = J_x(t) \hat i  + J_y(t) \hat j = 
 \frac{d}{dt} \langle x(t) \rangle \hat i +  
\frac{d}{dt} \langle y(t) \rangle \hat j.\label{Jxy}
\end{equation}
In Fig.\ref{x-cur} we show the average $x$-position of the particle 
as a function of time, the current $J_x(t)$ being 
the derivative of such a curve. The current along $y$ is zero,
$J_y(t) = 0$, as expected.
The current occurs, however, in an interesting manner. First, we
realize that the particle feels the effect of the ratchet only when it
is inside the filament. Outside, it performs diffusive
standard Brownian motion. The net result is that the current tends,
asymptotically and very slowly, towards zero. However, for any finite
amount of time, the current is different from zero. Thus, a net
transport is always realized. Another form of saying this is that, 
even though the particle may stay a long time outside the
filament, it effectively has directional motion since it eventually
returns to it. This is better seen in Fig.\ref{nubes}, where we
present four ``snapshots'' of the position probability distribution
$\rho(x,y,t)$ for different times, and obtained from 2000
realizations of the process.  The distribution acquires an arrow-like
structure because of the ratchet within the filament and the centroid
moves always to larger values of $x$, i.e. the distribution appears to
be dragged by the filament.
The shape of the position distribution $\rho(x,y,t)$ suggests also
a peculiar behavior of its second moment,
\begin{eqnarray}
\Delta r^2 &=&\Delta x^2 + \Delta y^2 \nonumber \\
&=& \left(\langle x^2 \rangle -\langle x\rangle^2\right)
+ \left(\langle y^2 \rangle -\langle y\rangle^2\right) .\label{Dxy}
\end{eqnarray}
In Fig. \ref{msd} we show these three deviations. The total second
moment $\Delta r^2$ follows a power law $t^\alpha$ with $\alpha\approx
1.5$ after $t \approx 100$ time units.  That is, the system
effectively presents superdiffusion, in contrast to the
one-dimensional case where it is always normal diffusion.  The super
diffusive behavior is also present in the $x$-direction, while in the
$y$ direction the diffusion is normal.

The explanation for this behavior appears to be simple: When the 
particle is bound to the filament it feels the effect of the ratchet
and tends to move preferentially in the positive $x$ direction. 
When it is outside of the channel performs a normal random walk, with
normal diffusion, but it does not move preferentially to any
direction. Thus, for those 
times when it is outside the filament, even though 
appears statistically to be dragged
by the filament as mentioned above, the particle lags behind the centroid,
hence effectively producing a wider distribution and a larger
second moment. It is interesting to point out that the superdiffusive
behavior is a consequence of ``standard'' noises coupled to a highly
nonlinear process and does not arise from an exotic stochastic
process\cite{Katja}.

\section{A ratchet mechanism behind protein folding?}

Following the classical work of Anfinsen\cite{Anfinsen} on protein folding, 
Levinthal\cite{Levinthal} argued that if the multidimensional
(free) energy landscape had a ``golf course'' like shape, with the hole being
the protein native state, it would take essentially forever for a protein
to fold correctly, provided the search were performed at random, namely,
by Brownian motion. Since proteins {\it in vivo} fold extremely fast and
efficiently\cite{Alberts} the concept of a {\it funneled}
energy landscape was developed\cite{Wolynes,Dill}, with the further possibility of
folding pathways in which the protein folds following a path
towards or within the funnel leading to the native state. But very 
importantly, the process being driven essentially by free energy differences,
that is, by ``falling'' into the state of lowest free energy.
This idea has been central in the study of protein folding and we 
do not pretend to review all the advances made by the many groups
working on this field\cite{groups} nor to point out any possible
pitfalls of the theories. Rather, we would like to point out
an alternative way out to the so-called Levinthal paradox, without
necessarily requiring a {\it funnel} leading to the native state. 
We shall appeal to a ratchet mechanism for the process of protein folding.  
This model requires, in addition to having a ratchet-like energy 
landscape as we discuss below,
that the folding is a non-equilibrium process being driven by an ``external''
source.

\begin{figure}[ht]
\centering
\scalebox{.75}{\includegraphics{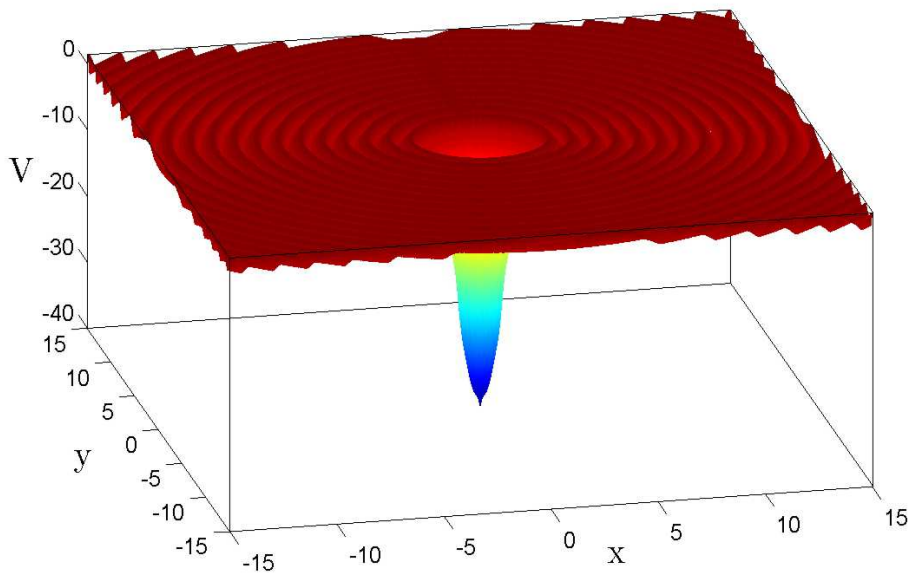}}
\caption{Two-dimensional ``free energy'' landscape of a protein,
see Eq.(\ref{ratch-prot}).
The landscape is flat on the average with a ratchet-like structure
on top of it. The deep well in the center is the ``native'' 
structure of the protein.}
\label{land-prot}
\end{figure}

The model consists of a ``particle'' moving in a multidimensional
space, the coordinates representing the different configurations of
a protein, or effective degrees of freedom\cite{note}. 
The particle is affected by a conservative ratchet potential
which maybe flat on the average (i.e. no need of a funnel) 
and with a ``small and deep enough hole'' representing
the native state. On top of the average potential there is an asymmetry
that makes the potential {\it toward} the native state, different
than in the opposite direction. Fig. \ref{land-prot} 
is a two-dimensional example for visualization, 
but we have realized many multidimensional potentials with the same
type of ``toward-away'' asymmetry, see below. Although we cannot
justify the existence of such a potential, one may argue that
the regularity specially in the secondary structures, at least at first order, 
suggest a kind of ``universal''
regular potential. Our proposal, undoubtedly not quite
justified, is that such a regularity may have a ratchet-like
form; after all, any coiling structure has an asymmetry since
it distinguishes between ``right'' and ``left''. The presence
of the thermal bath needs no justification and simply represents 
the effect of a viscous environment at a fixed temperature.
With only the last two ingredients, the ratchet potential and the bath,
the ``particle'' will obey Levinthal paradox and would not find
its way in a reasonable time, specially in a multidimensional space.
However, if the there is an {\it external} source of energy, with 
random properties and zero bias, acting on the particle, the latter
will definitely find its way toward the native state and,
on the average, could do so in a short time. 
The efficiency of the process, of course, will depend on the
values of the different parameters. What is the origin of the
external source? We may generically argue that processes in  
living organisms are through states of non-equilibrium with all sorts
of gradients of different physical properties,  and that the 
maintenance of those gradients may be traced back to the consumption
of ATP. In a more specific way, although it is not clear that
occurs in all proteins, it is known that many proteins fold
assisted by chaperone proteins, which in turn, consume 
ATP\cite{Ellis}. Thus, the proposed mechanism requires the presence
of an external source, which in our opinion, is physically appealing
since Life needs such a source for occurring as required by the
Second Law. Nonetheless, we cannot
say exactly what the mechanism is at the level of the protein.

The mathematical model is, therefore, a multidimensional version
of Eqs.(\ref{elx}) and (\ref{ely}) for the kinesin-on-a-microtubule
model, but now the ratchet potential has an overall (
topologically equivalent) spherical symmetry, with a ``towards-away''
asymmetry. For the numerical results we present below we consider
periodic potentials such as the following,
\begin{eqnarray}
V(x_1,x_2,\dots,x_d)& =& V_0 \left[\sin(2\pi r/\lambda)
+ a \sin(4\pi r/\lambda) + b\sin(6 \pi r/\lambda) \right] \nonumber \\
&& -V_1 e^{- c r^2} .
\label{ratch-prot}
\end{eqnarray}
where $d$ is the dimension of the space, namely, the number
of effective degrees of freedom of the protein; 
$r = (x_1^2 + x_2^2 + \cdots + x_d^2)^{1/2}$; and $a$, $b$ , $c$,
$V_0$ and $V_1$ real positive numbers. The last term in Eq.(\ref{ratch-prot})
represents the ``hole'' of the native state. We want to insist
that the ratchet potential need not be periodic, that is, 
the saw-tooth structure can have ``quenched'' random distances between 
maxima and minima, such as in the one-dimensional
case, see Fig. 2. Furthermore, the landscape does not have to be flat,
namely, it can have valleys and peaks, wider and higher or lower than 
the average peaks of the ratchet, and the particle can still climb over the
peaks provided they are not above the stall load of the 
ratchet\cite{Parrondo}. Full analysis of these cases are beyond
the briefness of this report and it will be presented 
elsewhere.

\begin{figure}[ht]
\centering
\scalebox{.50}{\includegraphics{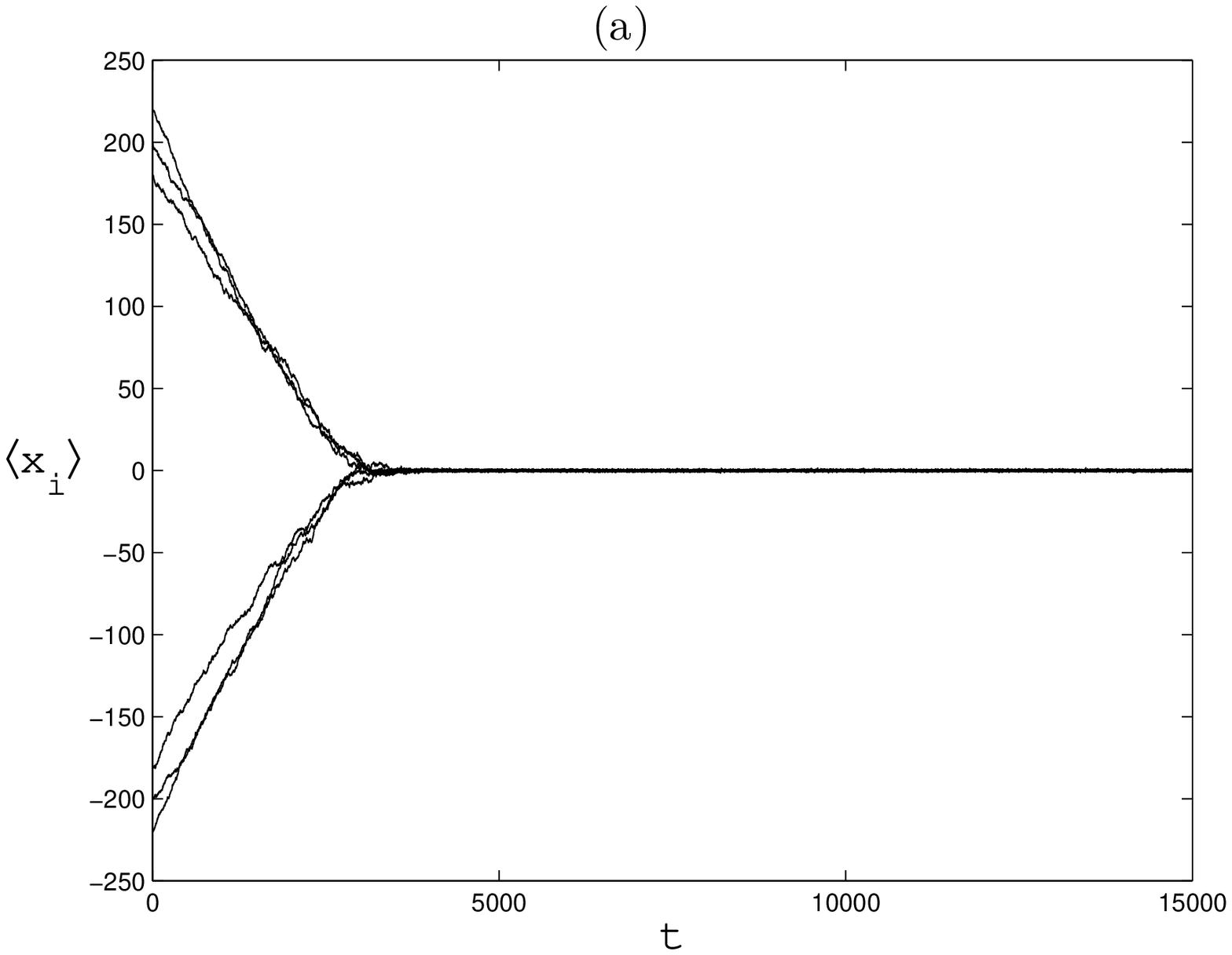}}
\scalebox{.50}{\includegraphics{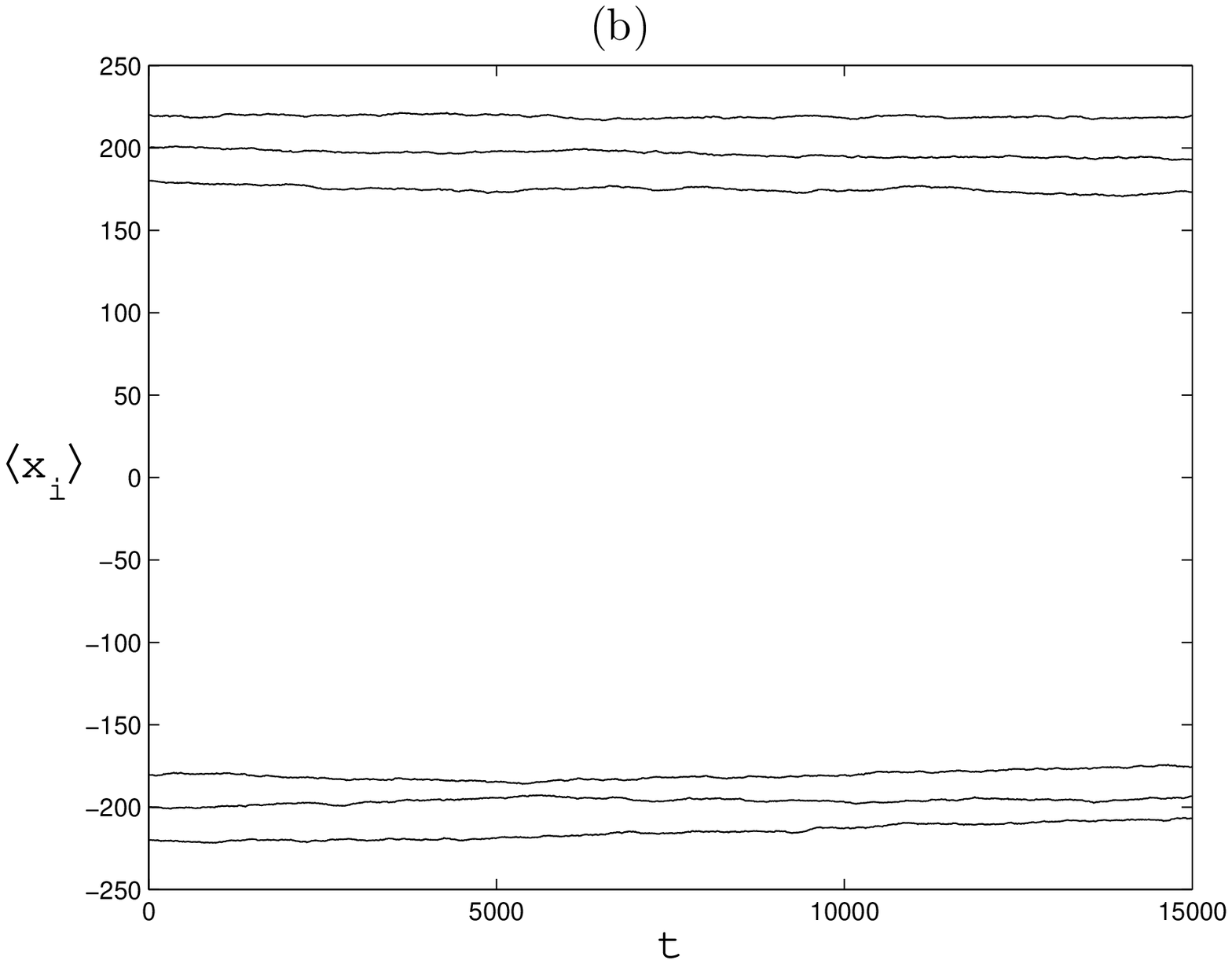}}
\caption{Average particle position 
$\langle x_i(t) \rangle$ in a six-dimensional space, $i=1,\dots,6$, 
as a function of time
for 100 realizations and same initial conditions. The spatial
period of the ratchet is $\lambda = 1$. (a) In the presence
of the external time dependent force, the ``protein'' folds. (b)
In the absence of the external time dependent force, the ``protein'' 
essentially never folds.}
\label{fold}
\end{figure}

Fig.\ref{fold} shows a typical example of the many different cases
we have analyzed. It corresponds to a ``protein'' with 6 degrees of
freedom. We show the average evolution of the protein 
coordinates $x_i(t)$, with $i = 1,2,\dots
,6$ over 100 runs for the same initial condition. In the presence
of the external force, and for those parameters, we found more than 95$\%$
``folding'', namely, the particle reached the ``native'' state in the
arbitrary maximum time of 15000 units.
The average time of folding can be read off the graph and corresponds
to 4000 unit times approximately, very short even in CPU time scales. 
When the external field was turned off, needles to say, 
we found 0$\%$ ``folding'', and we believe it will essentially 
never find its way no matter how long we run the program.

\section{Remarks} 

In this article we have revisited overdamped ratchet systems in
one dimension and extended it to several dimensions in order to
study biology related systems. For the case of one dimension we 
analyzed a novel {\it disordered} ratchet for the purpose of 
making explicit the fact that the space in which the systems
move do not correspond to a periodic variable, as it is 
routinely done in the literature. For these systems, strictly
speaking, do not exist a stationary state. The main result
that we want to stress here is the fact that, in the absence of
external time dependent forces, stochastic or deterministic,
there cannot be a current for all times and not only as an 
asymptotic condition. This is a requirement of the Second Law.
Once there is an external time dependent force, and the system
shows an intrinsic asymmetry, such as a ``left-right'' one,
there is nothing to prevent a current. The Second Law now acts
by setting limits for  the amount of work released, namely
on the efficiency of the process. We did not
address the latter limits although several workers have
discussed this point\cite{Parrondo}.

The ratchet mechanism has become a potential candidate
for many biological process, since, on the one hand
resembles a thermal engine at a mesoscopic level, 
but on the other it does 
not appear to need a specific design such as the man-made
engines. That is, by its inherent capacity to deliver work in 
the form of directional motion, one may be prompted to use
it as a model in somewhat obvious biological situations in which
a transport process, of some kind, is present. This has been
the case of the motion of kinesin proteins on microtubules, where
the protein appears to truly ``walk'' using ATP as the 
necessary input energy; there are other cases where ratchets have
been used as models of intracellular motion, or as molecular
motors, such as in 
the mitosis of the cell\cite{Jose}. 
Granted, all the models are
still at a very premature level of comparison with actual situations,
 mainly because the complicated biochemical
processes involved in real life can hardly be thought to be described 
by systems with one
or two degrees of freedom. Nevertheless, we believe it is worthwhile
to keep exploring these models not only by its potential relevance
in biological systems but for their own sake. Here, we have presented 
a two dimensional model of a kinesin in the presence of a 
quasi one-dimensional microtubule. The typical walks of the particle, 
wandering around until they hit the microtubule and then
directionally moving, show a striking similarity with actual kinesin 
motion, as seen in videos of these systems\cite{Alberts-videos}.
Although one can always try to fit these simple results to known
conditions, such as speed of kinesin walks or the energy
released by the consumption of ATP, we believe further research
is needed in two aspects of the model: first, we must confidently know 
how to relate the ratchet potential with the actual periodic structure
of the microtubule and the form in which the protein attaches to it, and
second, how to describe the consumption
of ATP by means of an appropriate statistical process.

The third section of this article on considering protein folding as 
being driven by a ratchet mechanism is highly speculative, but we
believe it is worthwhile to pursue it because, again, protein
folding appears as a process with directional motion. Traditionally,
this process has been thought to be driven by free energy differences
leading towards a minimum, similarly to a chemical reaction\cite{Wolynes},
and thus the idea of a funnel in the free energy landscape of the protein.
The presence of catalyzers and/or chaperone proteins may 
accelerate this process and their presence is welcome in the theory.
Here, by following the idea that Life processes consume energy
to yield their products, similarly to a man-made thermal engine, we have
speculated that protein folding is driven by an ``engine'' in which
the protein itself is part of it. This now makes a 
necessity the presence of choperones and/or catalyzers that in turn
consume energy. At this moment, the hardest part to justify in our
model is the ratchet structure of the energy landscape, and the best
that we can say is that it is motivating to see that 
the secondary structure of proteins is universally made of $\alpha$-helices
and $\beta$-sheets, structures with certain periodicity and asymmetry.
On the other hand, we do not have evidence neither pro nor con
that a ratchet structure is present since this would have to
be seen in the highly multidimensional energy landscape. As mentioned
before, the ``folding'' with this mechanism does not need a funnel, but 
of course, the interplay of both properties would make the process
even more efficient. The simulations that we have performed, so far 
up to a landscape in 6 dimensions, are extremely encouraging since
they can easily be tuned to a very fast and almost 100 $\%$ ``folding''.
Extending to an arbitrary number of dimensions only requires longer
computer time and, at present, does not add anything fundamentally 
different. What is really needed is the actual form of an energy landscape,
a hard problem with many researchers very much interested in it.
However, without yet knowing the actual form of an energy landscape
in its many dimensions, we can make models in which the ratchet is not
periodic and in which the landscape have valleys and peaks to test
the efficiency of the search for the ``native'' state. We shall present
those results in future contributions.


\begin{thebibliography}{99}

\bibitem{Magnasco} M. Magnasco, Phys. Rev. Lett. 71 (1993) 1477. 

\bibitem{Reimann} P. Reimann, Phys. Rep. 2 (2001) 237, and
references therein.

\bibitem{Hanggi1} P. Hanggi, F. Marchesoni, and F. Nori,
Ann. Phys. 14 (2005) 51, and references therein.

\bibitem{Astumian1} R.D. Astumian and P. Hanggi, Phys. Today
55 (2002) 33. 

\bibitem{vanKampen} N.G. van Kampen, {\it Stochastic Processes in 
Physics and Chemistry}, North Holland, Amsterdam, 1981.

\bibitem{Risken} H. Risken, {\it The Fokker-Planck equation}, Springer,
Berlin, 1984.

\bibitem{Feynman} R.P. Feynman, R.B. Leighton, and M. Sands,
{\it The Feynman Lectures on Physics}, Vol. I, Addison Wesley,
Reading, 1963.

\bibitem{Doering} C.R. Doering, W. Horsthemke, and J. Riordan,
Phys. Rev. Lett. 72 (1994) 2984.


\bibitem{Bartussek} R. Bartussek, P. Reimann, and P. Hanggi,
Phys. Rev. Lett. 76 (1996) 1166.

\bibitem{Astumian} R.D. Astumian and M. Bier, Phys. Rev. Lett.
72 (1994) 1766.

\bibitem{Mateos} see J.L. Mateos in this issue.

\bibitem{Alberts} B. Alberts, A. Johnson, J. Lewis, M. Raff,
K. Roberts, and P. Walter, {\it The molecular biology of the cell},
Garland, New York, (2002).

\bibitem{Wolynes} J.D. Bryngelson, J.N. Onuchic, N.D. Socci, and
P.G. Wolynes, Proteins: Struct., Funct., Genet. 21 (1995) 167.

\bibitem{Dill} K. Dill, S. Bromberg, K. Yue, K.M. Fiebig,
D.P. Yee, P.D. Thomas, and H.S. Chan, Protein Sci. 4 (1995) 561.
 
\bibitem{RR1} L.A. Ibarra-Bracamontes and V. Romero-Rochin,
Phys. Rev. E 56 (1997) 4048.

\bibitem{RR2} L. Viana and V. Romero-Rochin, Physica D 168 (2002) 193.

\bibitem{Parrondo} J.M.R. Parrondo and B.J. de Cisneros, Appl. Phys.
A 75 (2002) 179.

\bibitem{Fermi} E. Fermi, {\it Thermodynamics}, Dover, New York, 1956.

\bibitem{Ajdari} F. Julicher, A. Ajdari, J.P. Prost, 
Rev. Mod. Phys. 69 (1997) 1269.


\bibitem{Wang} Z. Wang, S. Khan, M.P. Sheetz, Biophys. J. 69 (1995) 2011.

\bibitem{Cilla} S.Cilla and L.M. Flor\'{\i}a, Il Nuovo Cimento, 20 
(1998) 1761.


\bibitem{Zhao} T.J. Zhao, Y.Z. Zhuo, Y. Zhan, Q. Yi, T.G. Cao,
Mod. Phys. Lett. B 16 (2002) 999.

\bibitem{Middleton} J. Middleton, J. A. Tuszynski, CRM Proceedings 
and Lecture Notes. 
39 (2004) 251.

\bibitem{Kostur} M. Kostur, L. Schimansky-Geier, Phys. Lett. A 265 (2000) 337.

\bibitem{Bao} J.D. Bao  Phys. Rev. E 62 (2000) 4606.

\bibitem{Bao2} J.D. Bao  Phys. Rev. E 63 (2001) 061112.

\bibitem{Eichhorn} R. Eichhorn, P. Reimann, P.Hanggi, Physica A
325 (2003) 101.

\bibitem{Lipowsky} R. Lipowsky, S. Klumpp, T. M. Nieuwenhuizen,
Phys. Rev. Lett. 87 (2001) 108101.

\bibitem{Nieu} T. M. Nieuwenhuizen, S. Klumpp, R. Lipowsky, 58 (2002)
468.

\bibitem{Katja} J.M. Sancho, A.M. Lacasta, K. Lindenberg, I.M. Sokolov,
and A.H. Romero, Phys. Rev. Lett. 92 (2004) 250601.

\bibitem{Anfinsen} C.B. Anfinsen, E. Haber, M. Sela, and F.H. White,
Proc. Natl. Acad. Sci. 47 (1961) 1309.

\bibitem{Levinthal} As quoted in Ref.\cite{Wolynes}.

\bibitem{groups} The number of researchers working on protein folding is
enormous, see the web page 
http://www.fccc.edu/research/labs/roder/folding$\_$groups.html.

\bibitem{Ellis} R.J. Ellis and F.-U. Hartl, FASEB J. 10 (1996) 20.

\bibitem{Jose} F. Gibbons, J.-F. Chauwin, M. Desp\'osito, and J.V. Jos\'e,
Biophys. J. 80 (2001) 2515.

\bibitem{note} The multidimensional free energy landscape may already have
the information that the system is immersed in water. That is, the
interaction potential should have all effective interactions due to
the hydrophilic or hydrophobic properties of the different aminoacids.


\bibitem{Alberts-videos} See the videos in the DVD disk accompanying
the book by Alberts et al., Ref.\cite{Alberts}.

\end{thebibliography}
\end{document}